\documentclass[journal,10pt,onecolumn,twoside]{IEEEtran}

\usepackage{cite}
\usepackage{amsmath,amssymb,amsfonts}
\usepackage{algorithmic,mathcomSTEv4}
\usepackage{graphicx}
\usepackage{textcomp}
\usepackage{xcolor}
\usepackage[]{algorithm2e}
\usepackage{setspace}
\newtheorem{theorem}{{\textbf{Theorem}}}

\newtheorem{remark}{{{Remark}}}

\begin{document}
\title{Wireless Federated Learning with Limited Communication and Differential Privacy}
\author{Amir~Sonee,
        Stefano~Rini and
        Yu-Chih Huang
\doublespacing\doublespacing
\thanks{Stefano Rini and Yu-Chih Huang are with the Department of Electrical Engineering and Computer Science,
National Yang Ming-Chiao Tung University (NYCU), Hsinchu, Taiwan.
Email: amir.sonee@mail.um.ac.ir,\{stefano,jerryhuang\}@nctu.edu.tw.
}
}



\maketitle
\vspace{-15mm}
\begin{abstract}
This paper investigates the role of dimensionality reduction in efficient communication and differential privacy (DP) of the local datasets at the remote users for over-the-air computation (\emph{AirComp})-based federated learning (FL) model.
More precisely, we consider the FL setting in which clients are prompted to train a machine learning model by simultaneous channel-aware and limited communications with a parameter server (PS)
over a Gaussian multiple-access channel (GMAC), so that transmissions sum coherently at the PS globally aware of the channel coefficients.
%
%
For this setting, an algorithm is proposed based on applying (i) federated stochastic gradient descent (FedSGD) for training the minimum of a given loss function based on the local gradients, (ii) Johnson-Lindenstrauss (JL) random projection for reducing the dimension of the local updates and (iii) artificial noise to further aid user's privacy.
%
%
%
For this scheme, our results show that the local DP performance is mainly improved due to injecting noise of greater variance on each dimension while keeping the sensitivity of the projected vectors unchanged. This is while the convergence rate is slowed down compared to the case without dimensionality reduction. As the performance outweighs for the slower convergence, the trade-off between privacy and convergence is higher but is shown to lessen in high-dimensional regime yielding almost the same trade-off with much less communication cost.
\end{abstract}

\vspace{-2mm}
\section{Introduction}
Recently, FL has emerged as a promising paradigm for distributed \emph{edge} learning over centralized networks focusing on \emph{edge} computations without the need to communicate users' large datasets. This provides capability of preserving privacy for the users' datasets as well as communication-efficiency.
This setting is relevant in a host of  modern-day training scenarios in which some deep learning model is to be trained over \emph{big data} available at a set of remote users whose privacy and anonymity has to be preserved in the course of learning process.
When remote users and PS are connected wirelessly, one can exploit the properties of the radio environment for broadband over-the-air model aggregation which greatly  reduces the communication latency while increasing bandwidth efficiency \cite{BAA_FEEL}.
In the current big data-intensive applications, the model exchanged between large number of remote users and the PS through training is relatively large, so that dimensionality reduction techniques can enormously facilitate computation, storage and communication over bandwidth-limited channels.
%
%
\subsubsection*{Literature Review}
Various approaches have been proposed in the literature to address FL performance in terms of communication efficiency, privacy, and \emph{AirComp}.
%
%
The efficiency schemes put forth in the literature mainly fall into two categories: gradient sparsification and gradient quantization.
Sparsification methods highly rely on fixed or variable rate elimination of the dimensions of the gradient vector based on a specific criterion such as magnitude or variance \cite{fixed_sparsification_FL,Atomo_FL,Alistarh2018Spars_FL,rTop-k-sparse}. This is while quantization methods focus on discretizing the gradient vectors through dimension-wise \cite{FL_DSGD_binomial} or vector quantization \cite{gandikota2019vqsgd}.
%
%
Data privacy in FL model has been mainly addressed through DP as a context-free notion
%
evaluating the privacy loss incurred by membership attacks to extract information about the individual sample points \cite{inferenceattack_FL}.
%
One most common method to preserve privacy is via local perturbation of the gradients by an artificial noise of Gaussian or Laplacian distributions \cite{DP_book,FL_wireless_LDP,wireless-DP-FL-adaptive,FL_DP_analysis}.
%
%
%
%
Finally, motivated by the use of FL in emerging technologies such as IoT, V2V and D2D communications between mobile or wireless \emph{edge} devices over wireless media, the principle of \emph{AirComp} has been put forth to further extend the original FL formulation presented for noiseless, dimension-unlimited channel to the \emph{AirComp} FL model
incorporating characteristics of the wireless radio environment in communication channel model by considering that transmissions between the clients and the PS occur over MAC \cite{BAA_FEEL,fadingwireless_FL}.
%
%
%
\subsubsection*{Contributions}
This paper leverages dimensionality-reduction technique featuring its further contribution to enhance privacy in FL setting in addition to efficiency.
Specifically, we propose a scheme referred to as \emph{differentially private random projection FedSGD} (\emph{DPRP-FedSGD}) addressing the interplay of these three ingredients
%
 %
in the FL problem formulation:  (i)  efficiency, (ii) privacy, and (iii) \emph{AirComp}.
We will show, in particular, that through appropriate use of a dimensionality-reduction linear random projection of JL type like Gaussian or Sub-Gaussian distributions that preserves almost isometry of the projected vectors, we can incorporate these issues through
%
(i) reducing the communication length for efficiency
(ii) bringing about more per-dimension Gaussian artificial noise with fixed noise power at clients for local DP (LDP), and (iii) inverting the aggregated vector through the transpose of the random projection matrix for \emph{AirComp}-aided update of \emph{global} model. Finally, we analyze the training and privacy performance of \emph{DPRP-FedSGD} in terms of the convergence rate and LDP of the underlying mechanisms and show that LDP is scaled down as $\mathcal{O}(\sqrt{r/d})$ while convergence is scaled up as $\Ocal(d/r)$. Moreover, an algorithm resulting in optimal convergence of \emph{DPRP-FedSGD} is proposed exploiting the static optimal noise power allocation and reduced dimension.
\subsubsection*{Notation}
$[n]$ represents the set of integers $\{1,\ldots,n\}$ and
$\|\xv\|_p$ indicates the  $\ell_p$-norm of vector $\xv$.
%
%
%

%
\section{Preliminaries}
\label{sec:Preliminaries}
\subsection{Federated Stochastic Gradient Descent (FedSGD)}
\label{sec:FL model}
As a distributed ML model, FL consists of $n$ clients aiming at collaborative optimization of an \emph{empirical loss} function
\ea{\label{eq:empirical-loss-function}
{\rm L}(\wv)=\f{1}{\left|\Dcal\right|}\sum_{i\in[n]}\left|\Dcal_{i}\right|{\rm L}_{i}\lb\wv\rb,
}
over the model vector $\wv\in\Rbb^d$ and under the coordination of the PS where ${\rm L}_i(\wv)$ is the \emph{local loss} function computed over the disjoint \emph{local} datasets $\Dcal_i$ at client $i$ with $\Dcal=\cup_{i\in[n]}\Dcal_i$.
The prevalent approach for numerical optimization of \eqref{eq:empirical-loss-function} is through iterative application of (synchronous) distributed stochastic gradient descent, also known as federated SGD (\emph{FedSGD}) over $T$ iterations. This is a large-scale variant of \emph{SGD} wherein each client $i$ locally computes the stochastic gradient vector at iteration $t\in[T]$, $t\neq1$, as $\gv_{i}^{t}
=\nabla{\rm L}_{i}\lb\wv^{t-1}\rb$ with access to the \emph{global} model update $\wv^{t-1}$ of previous iteration. Subsequently, the PS aggregates the \emph{local} gradients so as to obtain an unbiased estimation (stochastic gradient) of the true \emph{global} gradient $\nabla{\rm L}(\wv^t)$ as $\gv^{t}=\sum_{i\in[n]}\gv_{i}^{t}/n$ which is employed in \emph{global} model updating as $\wv^{t}=\wv^{t-1}-\eta^{t}\gv^{t}$ where $\eta^t$ is the iteration-dependant \emph{learning rate}.
A vector $\gv^{t}$ is called a stochastic gradient of ${\rm L}$ if ${\sf E}\lsb \gv^{t}\rsb=\nabla{\rm L}\lb\wv^{t}\rb$.


\subsection{Dimensionality-reduction via Random Projection}
\label{sec:dim-reduc}
Reducing the dimension of the transmitted local gradients (models) is generally considered as a \emph{sparsification} method. 
%
%
A suitable way for this approach is through the ubiquitous \emph{database-friendly random projection (RP)} as proposed by Johnson-Lindenstrauss (JL) \cite{JL}. The main idea comprises first generating a $d\times d$ random projection matrix (RPM) $\Uv$ with entries
drawn i.i.d. from a specific distribution satisfying the asymptotic orthogonality of the rows, and then projecting the space of $d$-dimensional vectors into the subspace of $r$ dimension using this random matrix as $\zv_i^t =\Uv\gv_i^t$.
Among the common distributions for random matrix projection, standard Gaussian or sub-Gaussian such as Achlioptas are widely used where the latter results in sparser random matrix projection.
The overall procedure of JL transformation leads to a high probability $\ell_2$-norm unbiased projection to a lower dimensional vector i.e. $|\|\zv_i^t\|_2^2/\|\gv_i^t\|_2^2-1|\leq\varepsilon$ with probability $1/n^a$ so long as $r\geq (4+2a)\lb \varepsilon^{2}/2-\varepsilon^3/3\rb^{-1}\ln n$, $0<\varepsilon<1$ and $a>0$ which we refer to as the JL  condition \cite{JL}.
\subsection{Differential Privacy}\label{sec:DP}
As the distributions of the databases at clients are unknown in the ML models, DP is perhaps the most rigorous context-free criterion to quantify and measure the privacy of the learning process.
%
%
%
When the PS is assumed to be \emph{curious but honest}, one suitable way to guarantee privacy is by  having clients individually apply randomized algorithm on their \emph{local} updates.
This latter approach is referred to as \emph{local differential privacy (LDP)}.
More specifically, let ${\rm f}_i:\Dcal_{i}\rightarrow\mathcal{Z}_i$ be the query function composed of providing the \emph{local} updates, based on the \emph{local} dataset  $\Dcal_i$, followed by an \emph{RPM} reducing the dimension,
then a mechanism ${\rm M}_i:\mathcal{Z}_i\rightarrow\mathcal{Y}$
releasing the output of the query function to the PS
is said to be $\lb\epsilon_i^t,\delta^t\rb$-LDP at client $i$ if for any $\zv_i^t,\zv_i^{\prime t}\in\Zcal_i\subseteq\Rbb^{r}$ and any measurable subset
$\Scal\subseteq\Ycal$,
\ea{\label{eq:DP-formula-def}
\textsf{Pr}\lsb{\rm M}_i\lb\zv_i^{\prime t }\rb\in\Scal\rsb\leq e^{\epsilon_i^t}\textsf{Pr}\lsb{\rm M}_i\lb\zv_i^t\rb\in\Scal\rsb+\delta^t.
}

The quantity $\epsilon_i^t$ can be equivalently viewed as the bound on \emph{privacy loss} $\mathfrak{L}_p^t=\ln\lb\textsf{Pr}\lsb{\rm M}_i\lb\zv_i^{\prime t}\rb\in\Scal\rsb/\textsf{Pr}\lsb{\rm M}_i\lb\zv_i^t\rb\in\Scal\rsb\rb$, attained with probability at least $1-\delta^t$  at client $i$ as ${\sf Pr}\lb|\mathfrak{L}_p^t|\leq\epsilon_i^t\rb\geq1-\delta^t$, measuring the indistinguishability between two sample points of its database given any observation subset of the mechanism output. Throughout this paper, we assume the output spaces $\Ycal,\Scal\subseteq\Rbb^{r}$.
In the context of FL, the privacy loss over $T$ iterations referred to as $T$-fold LDP is considered which is shown to guarantee the worst case $\lb \sum_{t\in[T]}\epsilon_i^t ,\sum_{t\in[T]}\delta^t\rb$-LDP by the \emph{composition theorem},
\cite{DP_book}.
One factor of paramount importance in limiting privacy loss is through bounding the change in some metric quantities of the output of the query function with the change in the input. This quantity is specified by the sensitivity of the query function which in case of a randomized query function is defined as follows. The query function ${\rm f}_i$ is said to be $\lb\Delta_{\ell_q},\delta^\prime\rb$ sensitive w.r. to $\ell_q$-norm if for any two neighbouring datasets $\Dcal_i$ and $\Dcal_i^\prime$ there exists two coupling random variables $Z_{i},Z_{i}^\prime\in\Zcal_i$ with the same marginal distribution as ${\rm f}_i\lb\Dcal_i\rb$ and ${\rm f}_i\lb\Dcal_i^\prime\rb$, respectively, such that ${\sf Pr}\lsb\left\|Z_i -Z_i^\prime\right\|_q\leq\Delta_q\rsb\geq1-\delta^\prime$, \cite{FL_DSGD_binomial}.

\subsection{MAC AirComp}
\label{sec:system_model}
In the following, we consider 
a form of \emph{AirComp} in which clients transmit their perturbed projected local gradients
simultaneously to the PS over a flat-fading MAC, described at each iteration by the input/output vector relationship  \vspace{-0.5mm}
 %
%
\ea{\label{eq:channel-model}
\yv^t=\sum_{i\in[n]}h_{i}^{t}\xv_i^{t}+\nv^t,\vspace{-1mm}
}
so that the aggregated gradients can be estimated from the channel output.
The channel coefficients $h_{i}^t$ are assumed to be constant over each iteration and known locally at each client and globally at the PS. $\nv^t$ is the  additive noise, assumed standard white Gaussian.
Also, the channel input $\xv_i^{t}$
is also subject to the average power constraint \vspace{-1mm}
\ea{\label{eq:power_constraint}
{\sf{E} }\lsb\left\|\xv_i^{t}\right\|_{2}^{2}\rsb\leq P_i.\vspace{-1mm}
}

Note  that it is assumed that
the down-link channel has infinite capacity
%
and for the PS to update the \emph{global} model based on an estimation of the true gradient, a post-processing operation on the received vector is carried out as $\hat{\gv}^t=f_r \lb\yv^t \rb$.

\section{Problem Formulation and Proposed Approach}
We consider a setting combining the three components of an FL in Sec. \ref{sec:Preliminaries}.
In particular, we assume that \emph{FedSGD} takes place in the setting in which one computation of the  gradient is sent by $r<d$ transmissions over the MAC in \eqref{eq:channel-model}.
%
%
For this scenario, we consider the problem of designing efficient communication algorithms which maximize the convergence rate of the model estimate to the optimal value under constraint on (i) the target $T$-fold LDP, and (ii) the communication taking place over $r$ channel uses as the MAC in \eqref{eq:channel-model}.
The convergence performance in terms of the optimality gap $\xi(T)$, defined as
\ea{
\xi(T)={\sf E}\left[{\rm L}\lb\wv_{T}\rb\right]-{\rm L}\lb\wv^{\ast}\rb,
\label{eq:optimaality}
}
where $\wv^{\ast}$ is the unique solution of the minimization of \eqref{eq:empirical-loss-function}.
Note that the expectation in \eqref{eq:optimaality} is over the randomness in the channel noise, as well as any source of randomness in the communication scheme.
In conclusion, for a given learning problem of dimension $d$, and with $L$-smooth (having $L$-Lipschitz continuous gradients)
and $\la$-strongly convex loss function, 
the performance in \eqref{eq:optimaality} is a function of the time horizon $T$ as well as (i) the number of channel transmissions for gradient update $r^t$, (ii) the coefficients of the MAC at each iteration $t$, and (iii) the  target privacy level $\ep_i^T$ at user $i$ at iteration $T$.
\subsection{DPRP-FedSGD Scheme}
\label{sec:Proposed Approach}
Based on the techniques in Sec. \ref{sec:Preliminaries}, we propose the following transmission strategy referred to as \emph{DPRP-FedSGD} Scheme.
%
%

\noindent
{\bf RPM construction:}
First, the random projection matrix $\Uv\in\Rbb^{d\times d}$ is generated with entries drawn independently
from Rademacher distribution (symmetric Bernolli taking values $+1$ and $-1$ with probability $1/2$), or
according to the Gaussian distribution of zero mean and unit variance as $\lsb\Uv\rsb_{i,j}\sim\Ncal\lb0,1\rb$, or Achlioptas distribution, $i,j\in[d]$, given by
\ea{
[\Uv]_{i,j}=\left\{\
\begin{array}{cc}
    +\sqrt{s}, &  1/2s\\
    0,  & 1-1/s\\
    -\sqrt{s}, & 1/2s
\end{array}
\right.,
}
and is assumed to be shared between the clients and the PS through a random seed at each iteration.

\noindent
{\bf Gradient projection:}
Each client $i\in[n]$ at iteration $t\in[T]$ projects the local gradient $\gv_i^t$ into the an $\ell_2$-norm unbiased random vector $\zv_i^t$ as
\ea{\label{eq:projected-local-gradient}
\zv_i^t 
=\f{1}{\sqrt{r^t}}\Dv_{{\rm r}}\Uv\gv_i^t=\Tv\gv_i^t
}
where 
$\Dv_{{\rm r}}$ is a $r^t\times d$ rectangular diagonal matrix i.e. $\lsb\Dv_{\rm r}\rsb_{i,i}=1$, $i\in[r^t]$, and $\lsb\Dv_{\rm r}\rsb_{i,j}=0$, $j\neq i$. Such random projection into $r$-dimensional subspace, preserves the unbiasedness of the Euclidean-norm as ${\sf{E}}\lsb\|\zv_i^t\|_2^2\rsb=\|\gv_i^t\|_2^2$.

\noindent
{\bf \emph{AirComp}:}
client $i$ transmits a phase-compensated noisy scaled variant of the projected vector satisfying the power constraint:
\ea{\label{eq:transmitted-signal}
\xv_i^t =e^{-j\varphi_i^t}\lb\f{\sqrt{\gamma_i^t P_i}}{L}\zv_i^t +\sqrt{\f{\zeta_i^t P_i}{r^t}}\mv_i^t\rb,
}
where $\|\gv_i^t\|_2\leq L$ is the second-order bound of the local gradient, and $\gamma_i^t$ and $\zeta_i^t$ represent the fraction of the power dedicated to the transmission of the projected signal and the artificial noise, respectively, with $\gamma_i^t +\zeta_i^t\leq1$.

As a result of the channel model in \eqref{eq:channel-model}, the PS receives
\ea{
\label{eq:received-signal-PS}
 \yv^t =\sum_{i\in[n]}\f{\sqrt{\gamma_i^t \kappa_i^t}}{L}\zv_i^t+\sum_{i\in[n]}\sqrt{\f{\zeta_i^t \kappa_i^t}{r^t}}\mv_i^t +\nv^t,
}
at iteration $t$ where $\kappa_i^t=P_i|h_i^t|^2$ is the individual signal to noise ratio ({\sf SNR}) of client $i$ at the PS, and then makes the following post-processing to estimate the \emph{global} gradient for model updating\vspace{-1mm}
\begin{IEEEeqnarray}{rcl}\label{eq:postprocessing-global-gradient}
\hat{\gv}^t =\f{1}{nc^t}\Tv^{\sf{T}}\yv^t&=&
\f{1}{nc^t}\sum_{i\in[n]}\f{\sqrt{\gamma_i^t\kappa_i^t}}{r^t L}\overline{\Uv}_{\rm r}\gv_i^t\\
&&+\f{1}{nc^t}\sum_{i\in[n]}\f{\sqrt{\zeta_i^t\kappa_i^t}}{r^t}\Uv_{\rm r}^{\sf T}\mv_i^t+\f{1}{nc^t\sqrt{r^t}}\Uv_{r^t}^{\sf{T}}\nv^t\nonumber\vspace{-1mm}
\end{IEEEeqnarray}
where $\Uv_{\rm r}=\Dv_{\rm r}\Uv=\lsb\Uv_{{\rm r},1}|\ldots|\Uv_{{\rm r},d}\rsb$ and $\overline{\Uv}_r=\Uv_r^{\sf T}\Uv_{\rm r}$ with ${\sf E}\lsb\overline{\Uv}_{\rm r}\rsb=r^t\Iv_d$ for the three distributions generating $\Uv$. As a result, for the \emph{global} gradient estimation to remain unbiased i.e. ${\sf{E}}[\hat{\gv}^t]=\gv^t$, it is essential to have $\sqrt{\gamma_i\kappa_i^t}/L=c$
for some constant $c$ satisfying $\gamma_i\leq1$, $\forall i\in [n]$.
This corresponds to the value $c^t=\sqrt{\kappa_{\min}^t}/L$ where $\kappa_{\min}^t=\min_{i\in[n]}\kappa_i^t$,
and the fraction of the power allocated to the transmission of the projected gradient can be obtained as $\gamma_i^t=\kappa_{\min}^t /\kappa_i^t$.

As a result of this post-processing by the PS, the estimated \emph{global} gradient can be written as
\begin{IEEEeqnarray}{rcl}\label{eq:global-gradient-estimation}
\!\!\!\!\!\!\!\!\hat{\gv}^t &=&\f{1}{n}\!\sum_{i\in[n]}\!\!\f{1}{r^t}\overline{\Uv}_{\rm r}\gv_i^t\!+\!\f{1}{nc}\!\sum_{i\in[n]}\!\!\f{\sqrt{\zeta_i\kappa_i^t}}{r^t}\Uv_{\rm r}^{\sf T}\mv_i^t+\!\f{1}{nc\sqrt{r^t}}\Uv_{\rm r}^{\sf{T}}\nv^t\vspace{-1mm}
\end{IEEEeqnarray}
where the first term corresponds to the true global gradient and the other two terms is the equivalent noise vector of dimension $d$ appearing as a result of \emph{AirComp}.

\section{Main Results}
\label{sec:Main Results}
In this section, we first present the performance of \emph{DPRP-FedSGD} algorithm in terms of the LDP analysis in Sec.~\ref{sec:LDP-analysis}. Specifically, in Theorem~\ref{LDP-result}, we rely on the JL lemma to show that for a given budget on the artificial noise power, LDP scales as $\Ocal\lb\sqrt{r(1+\varepsilon)/d}\rb$ providing a better privacy level with $r=\mathcal{O}(\ln n)$ compared to the case of no reduction with high gain in high dimension regime. Furthermore, in Theorem~\ref{LDP-no-JL}, the results for general $r$ is proved by invoking exponential concentration bounds. We then turn our focus to the convergence analysis in Sec.~\ref{sec:convergence-analysis} and show that the convergence scales almost as $\Ocal\lb d/r\rb$ introducing slower convergence compared with the no reduction case. Also, to achieve the same performance on the convergence bound (LDP) after a specific large number of iterations, LDP (convergence) performance remains almost the same for both schemes but with less communication cost for the dimensionality reduction case.
Based on the analysis in this section, numerical results in Sec.~\ref{sec:numerical} demonstrate that in high-dimensional regime and especially with high-level privacy , the \emph{DPRP-FedSGD} scheme allows us to find some operating points for $r$ releasing a very close performance in terms of the convergence-privacy trade-off compared to the non-dimensionality-reduction case.

\subsection{LDP analysis}
\label{sec:LDP-analysis}
Since post-processing performed by the PS to reconstruct the \emph{global} gradient does not affect the privacy mechanism based on \cite[Prop. 2.1]{DP_book}, it suffices to go through the channel output \eqref{eq:received-signal-PS} to investigate the LDP loss. As the equivalent noise of the signal received by the PS is Gaussian distributed, the local differential privacy loss at client $i$ can be upper bounded to $\epsilon_i$ with probability greater than $1-\delta$, $\delta\in[0,1]$, as
\ea{
\epsilon_i=\f{\Delta_\yv}{\sigma_{n_c}}\sqrt{2\ln\lb\f{1.25}{\delta}\rb},
}
where $\sigma_{\nv_c}^2$ is the variance of the effective noise at the output of the channel $\sigma_{\nv_c}^2=\sum_{i\in[n]}(\zeta_i^t\kappa_i^t/r^t)+1$,
and $\Delta_\yv$ is the high probability (that is, $\geq 1-\delta^\prime$), $\delta^\prime\in[0,1]$, $\ell_2$-norm sensitivity, \cite{FL_DSGD_binomial}, of the
query function ${\rm f}_i$ producing the projected vector at client $i$
as a randomized function of the local database $\Dcal_i$.
\begin{theorem}
\label{LDP-result}
The \emph{DPRP-FedSGD} scheme with an RPM of JL transformation type can guarantee $T$-fold  $\lb\epsilon_i^T ,\delta^T\rb=\lb \sum_{t\in[T]}\epsilon_i^t ,\sum_{t\in[T]}\delta^t+T/n^a\rb$-LDP where
\ea{\label{LDP-JL}
\epsilon_i^t
=2\sqrt{(1+\varepsilon)}\sqrt{\f{2\kappa_{\min}^t \ln\lb1.25/\delta^t\rb}
{\sum_{i\in[n]}\lb\zeta_i^t \kappa_i^t/r^t\rb+1}}
}
provided that the reduced dimension satisfies the JL condition $r\geq (4+2a)\lb \varepsilon^{2}/2-\varepsilon^3/3\rb^{-1}\ln n$, $0<\varepsilon<1$ and $a>0$.
\end{theorem}
\begin{IEEEproof}
It should be noted that if the reduced dimension satisfies the JL condition, then $\ell_2$-sensitivity of the projected vectors lies within $(1-\varepsilon)$ and $(1+\varepsilon)$ of the $\ell_2$-sensitivity of the \emph{local} gradient $\|\gv_i^t\|_2$ with probability $1-1/n^a$ regardless of the type of distribution adopted for RPM i.e. $(1-\varepsilon)\left\|\gv_i^t-\gv_i^{\prime t}\right\|_2^2\leq
\left\|\zv_i^t-\zv_i^{\prime t}\right\|_2^2 \leq (1+\varepsilon)\left\|\gv_i^t-\gv_i^{\prime t}\right\|_2^2$. Accordingly, as the RP mapping is $\lb\sqrt{1+\varepsilon}\left\|\gv_i^t-\gv_i^{\prime t}\right\|_2,1/n^a\rb$ sensitive and the Gaussian mechanism is $\lb \sum_{t\in[T]}\epsilon_i^t ,\sum_{t\in[T]}\delta^t\rb$-LDP, the composition is
$\lb\epsilon_i^T ,\delta^T\rb$-
LDP by \cite{FL_DSGD_binomial}.
\end{IEEEproof}

It should be noted that as we increase the precision of the sensitivity for the projected vector, almost isometry is achieved with probability close to one, i.e. $\Delta_\zv \stackrel{{\rm w. p. 1}}=\Delta_{\gv}\leq 2L$, using a universal linear RP of polynomial time and independent of the datasets and gradients. Moreover, in this case,
the LDP of the proposed scheme outperforms the one without dimensionality reduction given the same level of total power for the artificial noise vector. This is roughly expected as the sensitivity is preserved with high probability after JL transform while the amount of noise variance per dimension is increased and hence contribute more to privacy.

However, in case the reduced dimension is not satisfying the high-probability $\ell_2$-norm concentration of the projected vector as in JL condition, for any value of $a$ and $\varepsilon$, the following result can be derived regarding the LDP.
\begin{theorem}\label{LDP-no-JL}
The \emph{DPRP-FedSGD} scheme with an RPM generated according to Achlioptas distribution can guarantee $T$-fold $(\epsilon_i^T ,\delta^T)=( \sum_{t\in[T]}\epsilon_i^t ,\sum_{t\in[T]}\delta^t +T\delta^\prime)$-LDP where
\ea{\label{LDP-Achlioptas1}
\!\!\epsilon_i^t=2\sqrt{\lb\!1\!+\!8s\sqrt{\f{\ln(1/ \delta^\prime)}{r^t}}\rb}\sqrt{\f{ 2\kappa_{\min}^t\ln\lb1.25/\delta^t\rb}{\sum_{i\in[n]}\lb\zeta_i^t \kappa_i^t /r^t\rb +1}},
}
when $r^t\geq \ln(1/\delta^\prime)$ and
\ea{\label{LDP-Achlioptas2}
\epsilon_i^t=2\sqrt{\lb1+8s\f{\ln(1/\delta^\prime)}{r^t}\rb}\sqrt{\f{ 2\kappa_{\min}^t\ln\lb1.25/\delta^t\rb }{\sum_{i\in[n]}\lb\zeta_i^t \kappa_i^t /r^t\rb +1}},
}
when $r^t <\ln(1/\delta^\prime)$.
A similar result can be derived with $s=1$, in case of an RPM generated according to Rademacher or Gaussian distribution.
\end{theorem}
\begin{IEEEproof}
The proof consists of providing a tight high-probability $\ell_2$-sensitivity bound for part of the channel output corresponding to the transmitted signal of an individual client by invoking the tail bound for sub-exponential random variables. The details can be found in
Appendix \ref{sec:appendix A}.
\end{IEEEproof}
\begin{remark}
The per-iteration LDP result for the \emph{FedSGD} without dimensionality reduction was derived in \cite{FL_wireless_LDP} as
\ea{
\epsilon_i^t=2\sqrt{\f{2\kappa_{\min}^t\ln\!\lb1.25/\delta^t\rb}{\sum_{i\in[n]}\lb\beta_i^t \kappa_i^t/d\rb +1}},
}
where $\beta_i$ is the fraction of the power allocated to the artificial noise. As long as the reduced dimension $r$ satisfies
\begin{IEEEeqnarray}{rcl}
\!\!\!\!\!\!r^t\!\leq\!
\f{\lb\sum_{i\in[n]}\zeta_i^t \kappa_{i}^t\rb\!
\lb1+\sqrt{1+\f{\lb1+\sum_{i\in[n]}\lb\beta_i^t\kappa_i^t/d \rb\rb}{\lb\sum_{i\in[n]}\zeta_i^t \kappa_i^t\rb^2}}\rb^{-1}
}{\sum_{i\in[n]}\lb\zeta_i^t\kappa_i^t/d\rb +32s^2\ln(1/\delta^\prime)}
\end{IEEEeqnarray}
then the dimensionality reduction using RPM outperforms in terms of the LDP. Taking the JL transform into account then $r$ should satisfy
\ea{
r^t <
 \f{\sum_{i\in[n]}\zeta_i^t \kappa_i^t}
{(1+\varepsilon)\lb\sum_{i\in[n]}\beta_i^t\kappa_i^t /d  \rb+\varepsilon^2}. \nonumber
}
The per-iteration LDP can be further upper bounded as
\begin{IEEEeqnarray}{rcl}
\epsilon_i^t
\leq2\sqrt{\f{r^t(1+\varepsilon)}{n}}\sqrt{\f{2\kappa_{\min}^t \ln\lb1.25/\delta^t\rb }{\min_{i}\zeta_i^t\kappa_i^t}},
\end{IEEEeqnarray}
which compared to the case of no dimensionality reduction, is smaller by the ratio $\sqrt{d/r^t (1+\varepsilon)}$ for the same artificial noise allocations.
\end{remark}

\subsection{Convergence analysis}\label{sec:convergence-analysis}
Next, we present our result for the convergence rate of the \emph{FedSGD} algorithm considering an $L$-smooth and $\lambda$-strongly convex loss function ${\rm L}$.
\begin{theorem}
\label{th:convergence-analysis}
For an $L$-smooth and $\lambda$-strongly convex loss function, the convergence rate of the \emph{DPRP-FedSGD} algorithm with learning rate $\eta^t=1/\lambda t$ using Achlioptas RPM can be upper bounded as
\begin{IEEEeqnarray}{rcl}
\label{eq:convergence-rate-Achlioptas}
\xi(T) \leq\!
\f{2L}{\lambda^2 T^2}\!\lsb \!\sum_{t\in[T]}\!L^2\!\lb\!1\!+\!\f{d\!+\!s^t\!-2}{r^t}\!\rb\!+\!\f{d}{n^2 (c^t)^2}\!\lb\sum_{i\in[n]}\!\f{\zeta_i^t \kappa_i^t}{r^t }\!+\!1\rb\!\rsb\!,\nonumber
\end{IEEEeqnarray}
with the similar result for Rademacher RPM if $s^t=1$. Also, the convergence result for a Gaussian RPM is given by
\begin{IEEEeqnarray}{rcl}
\label{eq:convergence-rate-Achlioptas}
 \xi(T) \leq
\f{2L}{\lambda^2 T^2}\!\lsb \!\sum_{t\in[T]}\!L^2\lb\!1\!+\!\f{d+1}{r^t}\!\rb\!+\!\f{d}{n^2 (c^t)^2}\!\lb\sum_{i\in[n]}\!\f{\zeta_i^t \kappa_i^t}{r^t}\!+\!1\rb\!\rsb.\nonumber
\end{IEEEeqnarray}
\end{theorem}
\begin{IEEEproof}
The proof is provided in Appendix
\ref{sec:appendix B}.
\end{IEEEproof}
\begin{remark}
For further interpretation of the convergence, let us consider the same reduction in dimension as $r^t=r$ and the same channel coefficients over all iterations as $h_i^t=h_i$ and so $c^t=c$, indicating a static power allocation $\gamma_i^t=\gamma_i$ and $\zeta_i^t=\zeta_i$. Accordingly, the bound on the convergence rate can be simplified and further related to the $T$-fold LDP as
\begin{IEEEeqnarray}{rcl}
\label{eq:utility-privacy}
\!\!\!\!\!\xi(T)
\leq
\f{2L^3}{\lambda^2 T}\!\lsb 1\!+\!\f{d+s-2}{r}\rsb\!+\!\f{16dL^3\ln\!\lb\!\f{1.25}{\delta}\!\rb(1+\varepsilon)T}{\lambda^2 n^2(\epsilon_i^T)^2},
\end{IEEEeqnarray}
which shows that for a given number of iterations the upper bound is decreasing with the target $T$-fold LDP implying the utility-privacy trade-off. Moreover, the RHS of this bound is convex with respect to $T$ indicating that gap is bounded for a given level of total privacy. This has been also verified in \cite{FL_DP_analysis} for the \emph{FedAvg} algorithm without considering dimensionality-reduction.
Compared to the bound provided on the optimality gap of the \emph{FedSGD} without dimensionality reduction, \cite{FL_wireless_LDP}, as
\begin{IEEEeqnarray}{rcl}
\xi(T)
&\leq
\f{2L^3}{\lambda^2 T}+\f{16dL^3\ln\lb\f{1.25}{\delta}\rb T}{\lambda^2 n^2(\epsilon_i^T)^2}
\label{eq:final bound}
\end{IEEEeqnarray}
it can be verified that, for a fixed number of iterations and when the number of clients is large, the bound on the convergence for the proposed reduction scheme in \eqref{eq:final bound} differs from the non-reduction case in that the first term is scaled by the ratio of the $d/r$.
This  scaling has also been observed in \cite{Distributed_Sub-gradient} for the case of cyclic projection in \emph{FedAvg}.
However, in case of large dimensions, both schemes have the same convergence performance for a given target $T$-fold LDP
since in this regime,  the second term of \eqref{eq:final bound} dominates.
Note though that our scheme attains this rate  more efficiently with limited communication of $r$ instead of $d$ dimensions per client.
\end{remark}

Next, we present a strategy on the static noise power allocation at clients and the reduced dimension to achieve the optimal convergence rate subject to power and $T$-fold LDP constraints at each client. More specifically, we address the following mixed integer nonlinear programming (MINLP) problem:
\begin{IEEEeqnarray}{rcl}
&\min_{r,\{\zeta_i\}_{i\in[n]}}& \f{2L}{\lambda^2 T}\lsb L^2\lb1\!+\!\f{d\!+\!s\!-\!2}{r}\rb\!+\!\f{d}{n^2 c^2}\!\lb\sum_{i\in[n]}\f{\zeta_i \kappa_i}{r}+1\rb\rsb\nonumber\\
&{\rm s.t.}&\ \gamma_i +\zeta_i\leq 1,\ \forall\ i\in[n]\nonumber\\
&&2\sqrt{(1+\varepsilon)}\sqrt{\f{2\kappa_{\min}^t \ln\lb1.25/\delta\rb}
{\sum_{i\in[n]}\lb\zeta_i \kappa_i/r\rb+1}}\leq \f{\epsilon_i^T}{T}\ \forall\ i\in[n]\nonumber\\
&&r\geq (4+2a)\lb \varepsilon^{2}/2-\varepsilon^3/3\rb^{-1}\ln n.
\end{IEEEeqnarray}
\begin{theorem}\label{th:optimal-convergence}
For the \emph{DPRP-FedSGD} algorithm, the optimal bound on the convergence subject to a given per-client LDP level and power constraints is given by
\begin{IEEEeqnarray}{rcl}
\!\!\!\!\!\!\!\!|\xi(T)|\!\leq\!\f{2L}{\lambda^2 T}\!\!\lsb\! L^2\!\lb\!1\!+\!\f{d\!+\!s\!-\!2}{r^{\ast}}\rb\!\!+\!\f{d}{n^2 c^2}\!\lb\sum_{i\in[n]}\!\omega_i(r^{\ast})\!+\!1\!\rb\!\rsb
\end{IEEEeqnarray}
where $\omega_i(r^{\ast}) =\min\lb\f{\kappa_i}{r^{\ast}}(1-\gamma_i),\lsb\Omega-\sum_{k=1}^{i-1}\f{\zeta_k \kappa_k}{r^{\ast}}\rsb^{+}\rb$ and $\Omega=\max_{i\in[n]}(1+\varepsilon)\f{8\kappa_{\min} \ln\lb1.25/\delta\rb}
{(\epsilon_i^T /T)^2}-1$ and $r^{\ast}$ is the largest value such that $\omega_i(r^{\ast}+1)=0$, $\forall i\in[n]$. The optimal values for the noise allocation coefficients are $\zeta_i^{\ast}=r^{\ast}\omega_i(r^{\ast})/\kappa_i$.
\end{theorem}
\begin{IEEEproof}
The proof is provided in Appendix \ref{sec:appendix C}.
\end{IEEEproof}

\section{Numerical results}\label{sec:numerical}
In this section, we provide numerical results though evaluation of the proposed performance results based on a scenario with $n=1000$ clients trying to train a strongly-convex loss function of $\lambda=0.001$ over the model parameter of dimension $d=10000$ used in classifying MNIST images through $T=1000$ iterations. The  clients transmit their local gradient updates subject to the same power constraint $P_i=1$, $\forall\ i\in[n]$ and over a channel with coefficients drawn according to complex standard Gaussian distribution as $\Ccal\Ncal(0,1)$. The random projection is assumed to be performed with a matrix of Achlioptas entries with $s=1$ and $s=2$.
Also, it is assumed that the $T$-fold LDP of each client should hold with probability at least $0.9$ and so $\delta=\delta^{\prime}=5\times 10^{-5}$ per-iteration.

\begin{figure}[htbp]
\centerline{\includegraphics[width=0.45\textwidth]{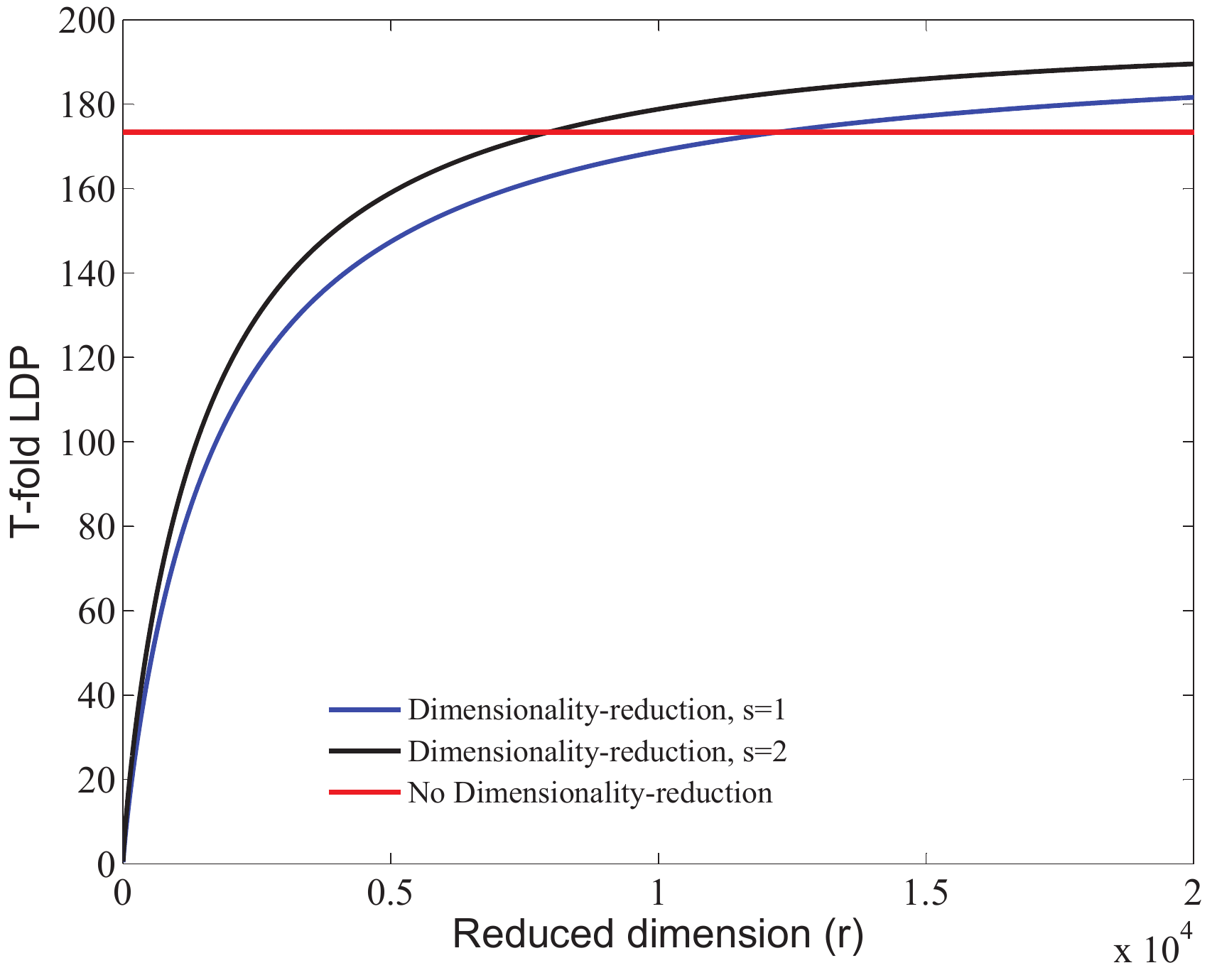}}
\vspace{-3mm}
\caption{$T$-fold LDP ($\epsilon_i^T$) versus reduced dimension compared with no dimensionality reduction, $T=1000$}
\vspace{-3mm}
\label{fig:LDP-DR}
\end{figure}

Fig.~\ref{fig:LDP-DR} shows that for a fixed budget on the power allocated on the artificial noise at clients, the \emph{DPRP-FedSGD} algorithm can surpass the scheme without dimensionality-reduction for a specific range of $r$, in terms of the $T$-fold LDP.

\begin{figure}[htbp]
\centerline{\includegraphics[width=0.45\textwidth]{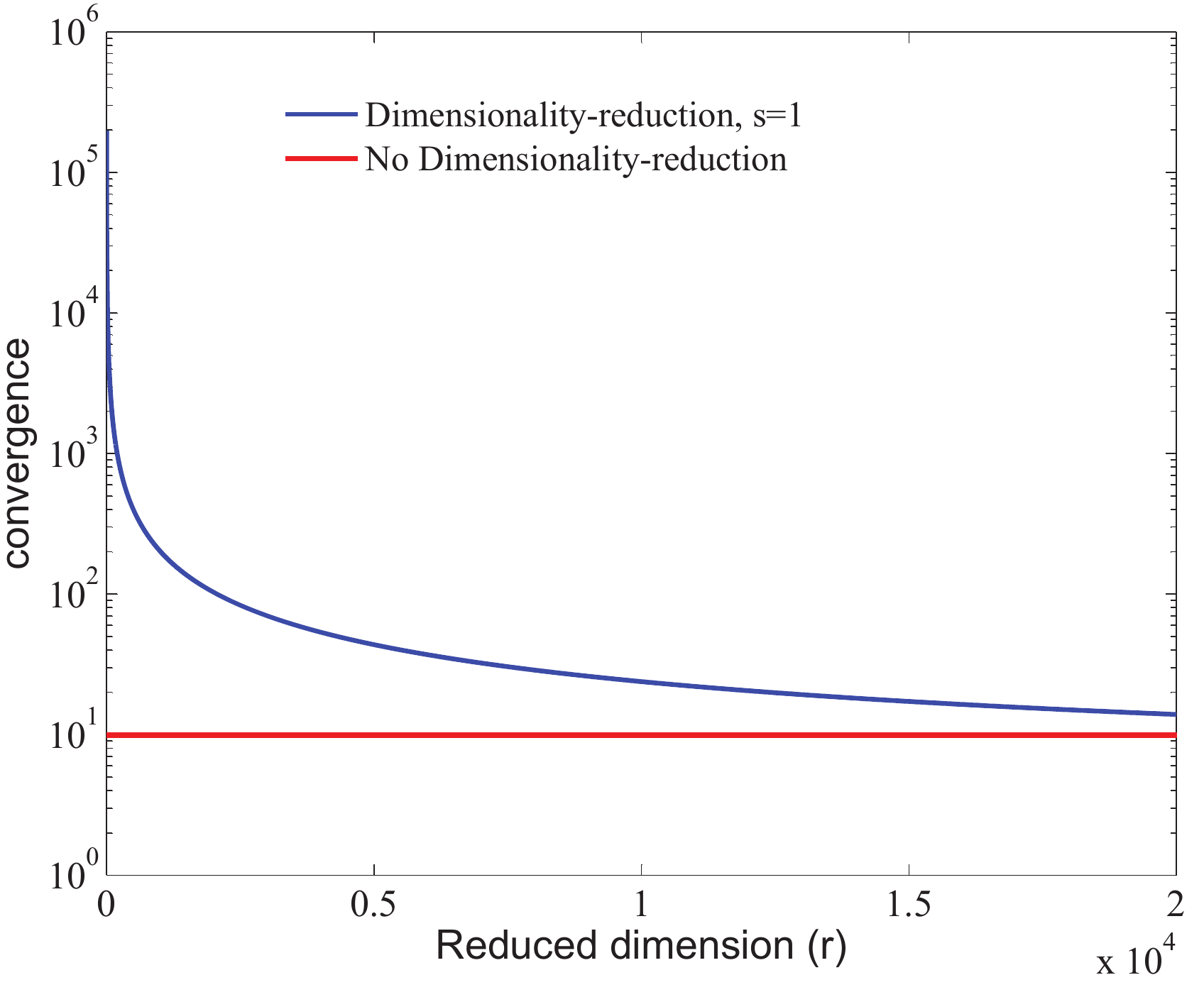}}
\vspace{-3mm}
\caption{Convergence bound versus reduced dimension compared with no dimensionality reduction, $T=1000$}
\vspace{-3mm}
\label{fig:conv-DR}
\end{figure}

\begin{figure}[htbp]
\centerline{\includegraphics[width=0.45\textwidth]{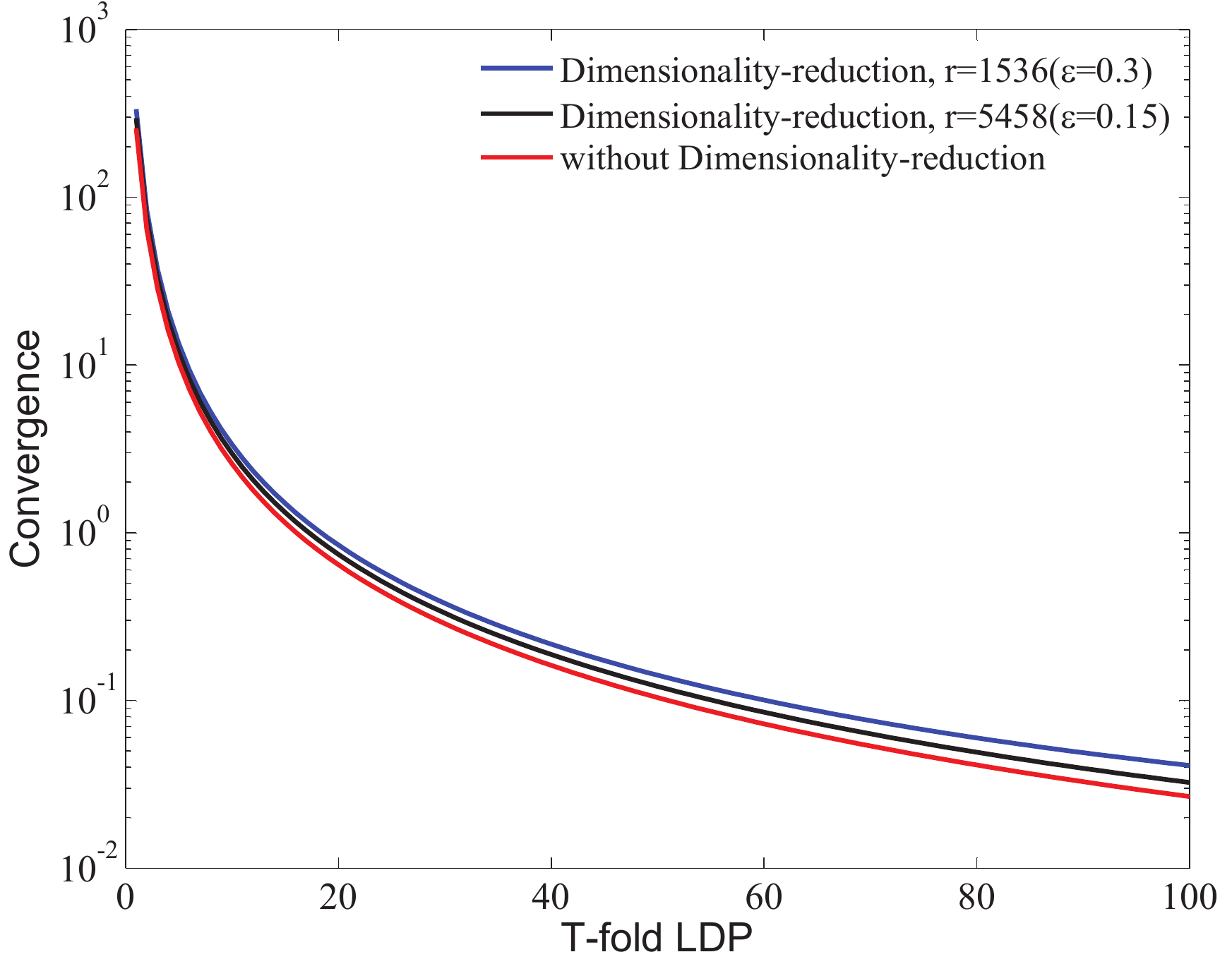}}
\vspace{-3mm}
\caption{Convergence bound versus $T$-fold LDP compared with no dimensionality reduction, $T=1000$}
\vspace{-3mm}
\label{fig:conv-LDP}
\end{figure}

In terms of the convergence, as shown in Fig.~\ref{fig:conv-DR}, the \emph{DPRP-FedSGD} scheme underperform the existing scheme that do not make use of reduction. This actually introduces that in terms of the trade-off between convergence and privacy, \emph{DPRP-FedSGD} presents lower performance but as in high-dimensional regime and specifically for stricter level of privacy, this trade-off, as shown in Fig.~\ref{fig:conv-LDP} is close to the corresponding performance without dimensionality-reduction.

\section{Conclusion}

In this paper, \emph{DPRP-FedSGD} scheme as a novel strategy to attain efficiency while preserving local differential privacy in  \emph{AirComp} federated learning was proposed and investigated.
More precisely, we considered random projection of JL transform for reducing the dimension of the local gradients at remote clients to $r<d$ with the aim of training the model through limited communication with the PS over a flat-fading MAC with much less channel uses than the one required to transmit the model size  $d$.
The projected gradients are then corrupted with artificial noise in order to enhance privacy and sent to the PS where they are accumulated and inverted by the transpose of the projection to update the \emph{global} model.
We provided an analysis on the differential privacy and convergence of the  \emph{DPRP-FedSGD} shows that under the same total artificial noise allocation, the LDP of clients outperforms the scheme in which only artificial noise is used for privacy by scaling down to $\Ocal(\sqrt{r/d})$. This is mainly a result of the projection making each dimension experience more noise while keeping the sensitivity of the projected vector almost unchanged. However, the optimality gap is scaled up by $\Ocal(d/r)$ resulting in slower convergence. This proposes a higher utility-privacy trade-off compared to the no projection scheme which can be almost mitigated in high-dimensional regime and hence guaranteeing almost the same performance with less communication cost.
\bibliographystyle{IEEEtran.bst}
\bibliography{GlobArxiv}
\appendices
\section{Proof of the Th. \ref{LDP-no-JL}}
\label{sec:appendix A}
we have to go through the high-probability $\ell_2$-sensitivity to further bound and compute the privacy loss. This can be formulated as the following tail bound for the  $\ell_2$-sensitivity random variable:
\begin{IEEEeqnarray}{rcl}\label{eq:sensitivity}
{\sf Pr}\!\lsb\f{\sqrt{\gamma_i^t \kappa_i^t}}{L\sqrt{r^t}}\left\|\Uv_{\rm r}\lb\gv_i^t-\gv_i^{\prime t}\rb\right\|_2
\!\geq\!\Delta_{\zv}\rsb\!\leq\!\delta^\prime.
\end{IEEEeqnarray}
or equivalently as
\begin{IEEEeqnarray}{rcl}\label{eq:sensitivity2}
{\sf Pr}\!\lsb\left\|\Uv_{\rm r}\lb\gv_i^t-\gv_i^{\prime t}\rb\right\|_2^2
\!\geq\!\f{r^t L^2\Delta^2_{\zv}}{\gamma_i^t \kappa_i^t}\rsb\!\leq\!\delta^{\prime}.
\end{IEEEeqnarray}

This result can be further tightened by approximating the square of the $\ell_2$-norm random variable $S=\left\|\Uv_{\rm r}\lb\gv_i^t-\gv_i^{\prime t}\rb\right\|_2^2$ through
a sub-exponential random variable. To show this, we should note that $S$ can be written as the sum of the squares of the inner products between each row $k\in[r^t]$ of the reduced matrix $\Uv_{\rm r}$ ($k^{\rm th}$ column of $\Uv_{\rm r}^{\sf T}$) and the vector $\gv_i^t -\gv_i^{\prime t}$ as
\ea{
S=\sum_{k\in[r]}\left<\Uv_{{\rm r},k}^{T},\gv_i^t -\gv_i^{\prime t}\right>^2=\sum_{k\in[r^t]}\lb\sum_{j\in[d]}\lsb\Uv\rsb_{k,j}\lsb\gv_i^t -\gv_i^{\prime t}\rsb_j\rb^2.
}

As the entries of the random projection matrix (RPM) are generated independently according to Rademacher, Achlioptas or Gaussian rv which are all Sub-Gaussian with parameter $\sigma_{U_{k,j}}$ (referred to as $\sim{\sf SubG}(\sigma_{U_{k,j}})$), then so is their linear combination with parameter
\ea{
\sigma_{U_{S_k}}=\sqrt{\sum_{j\in[d]}\sigma^2_{U_{k,j}}\lsb\gv_i^t -\gv_i^{\prime t}\rsb_j^2}=\sqrt{\sum_{j\in[d]}\sigma^2_{U}\lsb\gv_i^t -\gv_i^{\prime t}\rsb_j^2}=\sigma_{U}\left\|\gv_i^t -\gv_i^{\prime t}\right\|_2.
}
where since the entries are identically distributed $\sigma^2_{U_{k,j}}=\sigma_{U}$ with $\sigma_U=1$ for Rademacher and Gaussian distributions and $\sigma_U=s$ for Achlioptas distribution.

As the inner product is $\sim{\sf SubG(\sigma_{U_{S_k}})}$ then its square is Sub-exponential with parameters $(\sigma_S,b)$ (referred to as ${\sf SubE}(\sigma_S,b)$) where $\sigma_{S_k}=4\sqrt{2}\sigma_{U_{S_k}}^2$ and $b_{k}=4\sigma_{U_{S_k}}^2$, \cite{SubE-SquareSubG}. As a result, their sum is $\sim{\sf SubE\lb\sigma_S^\ast ,b^\ast\rb}$ where
\begin{IEEEeqnarray}{rcl}
\sigma_S^\ast&=&\sqrt{\sum_{k\in[r^t]}\sigma_{S_k}^2}=\sqrt{\sum_{k\in[r^t]}\sigma_{S}^2}=\sqrt{r}\sigma_{S}=4\sqrt{2r}\sigma^2_{U}\left\|\gv_i^t -\gv_i^{\prime t}\right\|_2^2\nonumber\\
b^{\ast}&=&\max_{k\in[r^t]}b_k =4\sigma_{S}^2=4\sigma^2_{U}\left\|\gv_i^t -\gv_i^{\prime t}\right\|_2^2
\end{IEEEeqnarray}

Now, we can propose the tail bounds for the $\ell_2$-sensitivity in \eqref{eq:sensitivity2} based on the tail bounds provided for sub-exponential rv with parameters $(\sigma_{S}^{\ast},b^{\ast})$ as
\ea{
{\sf Pr}\lsb S-\mu_S >\lambda\rsb\leq
\left\{
\begin{array}{cc}
    e^{-\lambda^2 /2\sigma_{S}^{\ast 2}} & 0\leq\lambda\leq \sigma_{S}^{\ast 2}/b^{\ast} \\
    e^{-\lambda/2b^{\ast}} & \lambda\geq \sigma_{S}^{\ast 2}/b
\end{array}.
\right.
}

Accordingly, the tail bound on $S$ can be given as
\begin{IEEEeqnarray}{rcl}
{\sf Pr}\!\lsb S
\geq\f{r^t L^2\Delta^2_{\zv}}{\gamma_i^t\kappa_i^t}\rsb
&=&{\sf Pr}\!\lsb S-\mu_S
\geq\f{r^t L^2\Delta^2_{\zv}}{\gamma_i^t \kappa_i^t}-r^t\left\|\gv_i^t-\gv_i^{\prime t}\right\|_2^2\rsb\nonumber\\
&\leq&
\left\{
\begin{array}{cc}
    \exp\lb-\f{(r^{t})^2\lb\f{ L^2\Delta^2_{\zv}}{\gamma_i^t \kappa_i^t}-\left\|\gv_i^t-\gv_i^{\prime t}\right\|_2^2\rb^2}{64r^t\sigma_{U}^4\left\|\gv_i^t-\gv_i^{\prime t}\right\|_2^4}\rb, & \Delta_{\zv}\leq \f{3\sqrt{\gamma_i^t\kappa_i^t }}{L}\left\|\gv_i^t-\gv_i^{\prime t}\right\|_2 \vspace{+4mm}\\
    \exp\lb-\f{r^t\lb\f{ L^2\Delta^2_{\zv}}{\gamma_i^t \kappa_i^t}-\left\|\gv_i^t-\gv_i^{\prime t}\right\|_2^2\rb}{8\sigma^2_{U}\left\|\gv_i^t-\gv_i^{\prime t}\right\|_2^2}\rb,  & \Delta_{\zv}>\f{3\sqrt{\gamma_i^t \kappa_i^t}}{L}\left\|\gv_i^t-\gv_i^{\prime t}\right\|_2
\end{array}
\right.\vspace{+1mm}
\nonumber\\
&\stackrel{(a)}\leq&
\left\{
\begin{array}{cc}
    \exp\lb-\f{r^t\lb\f{ L^2\Delta^2_{\zv}}{\gamma_i^t \kappa_i^t}-4\rb^2}{64\times16\sigma_{U}^4}\rb, & \Delta_{\zv}\leq 6\sqrt{\gamma_i^t\kappa_i^t}=6\sqrt{\kappa_{\min}^t} \vspace{+4mm}\\
    \exp\lb-\f{r^t\lb\f{ L^2\Delta^2_{\zv}}{\gamma_i^t \kappa_i^t}-4\rb}{32\sigma^2_{U}}\rb,  & \Delta_{\zv}>6\sqrt{\gamma_i^t \kappa_i^t}=6\sqrt{\kappa^t_{\min}}
\end{array}
\right.
\end{IEEEeqnarray}
where $(a)$ follows by the $L$-smooth condition of the loss function indicating that the gradient is $L$-Lipschitz continuous i.e. $\forall\ \wv,\wv^{\prime}\in\Rbb^{d}$,
$\left|\nabla{\rm L}\left(\wv^{\prime}\right)-\nabla{\rm L}\left(\wv\right)\right|\leq L\left\|\wv^{\prime}-\wv\right\|$.
This indeed implies that gradient and subgradients of the loss function are bounded i.e. $\|\nabla{\rm L}(\wv)\|_2\leq L$, $\forall\ \wv\in\Rbb^d$. Hence, $\left\|\gv_i^t -\gv_i^{\prime t}\right\|_2\leq 2L$ by the triangle inequality.

Assuming the {\bf Rademacher} and {\bf Gaussian} distributions for random matrix projection, then $\sigma_{U}=1$ and the high probability $\ell_2$-sensitivity holding with probability at least $1-\delta^\prime$ can be computed as
\begin{IEEEeqnarray}{rcl}\label{eq:sensitivity-Rademacher}
\Delta_{\zv}^t&=&
\left\{
\begin{array}{cc}
    2\sqrt{\kappa_{\min}^t}\sqrt{1+8\sqrt{\f{\ln(1/\delta^\prime)}{r^t}}}, & \Delta_{\zv}^t\leq 6\sqrt{\kappa_{\min}^t} \vspace{+5mm}\\
    2\sqrt{\kappa_{\min}^t}\sqrt{1+8\f{\ln(1/\delta^\prime)}{r^t}}, & \Delta_{\zv}^t>6\sqrt{\kappa_{\min}^t}
\end{array}
\right.
\end{IEEEeqnarray}
and so the LDP at client $i$ can be given as
\begin{IEEEeqnarray}{rcl}\label{eq:sensitivity-Rademacher}
\epsilon_i^t&=&
\left\{
\begin{array}{cc}
    2\sqrt{1+8\sqrt{\f{\ln(1/\delta^\prime)}{r^t}}}\sqrt{\f{2\kappa_{\min}^t\ln\lb1.25/\delta^t\rb}{\sum_{i\in[n]}\lb\zeta_i^t \kappa_i^t/r^t\rb +\sigma^2_{\nv}}}, & r^t\geq \ln(1/\delta^\prime) \vspace{+5mm}\\
    2\sqrt{1+8\f{\ln(1/\delta^\prime)}{r^t}}\sqrt{\f{2\kappa_{\min}^t \ln\lb1.25/\delta^t\rb}{\sum_{i\in[n]}\lb\zeta_i^t \kappa_i^t/r^t\rb +\sigma^2_{\nv}}}, & r^t< \ln(1/\delta^\prime)
\end{array}
\right. \vspace{-2mm}
\end{IEEEeqnarray}

Assuming the {\bf Achlioptas} distribution, then
then $\sigma_{U}=\sqrt{s}$ and the high probability $\ell_2$-sensitivity holding with probability at least $1-\delta^\prime$ can be computed as
\begin{IEEEeqnarray}{rcl}\label{eq:sensitivity-Rademacher}
\Delta_{\zv}^t&=&
\left\{
\begin{array}{cc}
    2\sqrt{\kappa_{\min}^t}\sqrt{1+8s\sqrt{\f{\ln(1/\delta^\prime)}{r^t}}}, & \Delta_{\zv}^t\leq 6s\sqrt{\kappa_{\min}^t} \vspace{+5mm}\\
    2\sqrt{\kappa_{\min}^t}\sqrt{1+8s\f{\ln(1/\delta^\prime)}{r^t}}, & \Delta_{\zv}^t>6s\sqrt{\kappa_{\min}^t}
\end{array}
\right.
\end{IEEEeqnarray}
and so the LDP at client $i$ can be given as
\begin{IEEEeqnarray}{rcl}\label{eq:sensitivity-Rademacher}
\epsilon_i^t&=&
\left\{
\begin{array}{cc}
    2\sqrt{1+8s\sqrt{\f{\ln(1/\delta^\prime)}{r^t}}}\sqrt{\f{2\kappa_{\min}^t\ln\lb1.25/\delta^t\rb }{\sum_{i\in[n]}\lb\zeta_i^t \kappa_{i}^t/r^t\rb +\sigma^2_{\nv}}}, & r^t\geq \ln(1/\delta^\prime) \vspace{+5mm}\\
    2\sqrt{1+8s\f{\ln(1/\delta^\prime)}{r^t}}\sqrt{\f{2\kappa_{\min}^t\ln\lb1.25/\delta^t\rb}{\sum_{i\in[n]}\lb\zeta_i^t \kappa_{i}^t/r^t\rb +\sigma^2_{\nv}}}, & r^t< \ln(1/\delta^\prime)
\end{array}
\right.
\end{IEEEeqnarray}

\section{Proof of Th. \ref{th:convergence-analysis}}
\label{sec:appendix B}
Considering the loss function ${\rm L}$ is $L$-smooth and $\lambda$-strongly convex
that is $\forall\ \wv,\wv^{\prime}\in\Rbb^{d}$,
\ea{
    {\rm L}\left(\wv^{\prime}\right)\geq {\rm L}(\wv)+\left<\gv,\left(\wv^{\prime}-\wv\right)\right>+\frac{\lambda}{2}\left\|\wv^{\prime}-\wv\right\|^2,
}
Then a formal analysis of the convergence rate for the \emph{FedSGD} algorithm is given by
\cite{GD_strong_convex},
\ea{\label{eq:convergence-FedSGD}
\left|\sf{E}\left[{\rm L}\lb\wv_{T}\rb\right]-{\rm L}\lb\wv^{\ast}\rb\right|\leq \f{2L}{\lambda^2 T^2}\sum_{t\in[T]}{\sf E}\lsb\|\hat{\gv}^t\|_2^2\rsb.
}
where
\ea{
\hat{\gv}^t =\underbrace{\f{1}{n}\sum_{i\in[n]}\!\f{1}{r}\overline{\Uv}_r\gv_i^t}_{\gtv_i^t}+\underbrace{\f{1}{nc}\sum_{i\in[n]}\!\f{\sqrt{\zeta_i\kappa_i^t}}{r}\Uv_r^{\sf T}\mv_i^t+\!\f{1}{nc\sqrt{r}}\Uv_{r}^{\sf{T}}\nv^t}_{\nv_{e}^t}
}
Accordingly,
we have to bound the second-order moment of the estimated \emph{global} gradient as \begin{IEEEeqnarray}{rcl}\label{eq:convergence-random-projection}
{\sf E}\lsb\|\hat{\gv}^t\|_2^2\rsb&=&{\sf E}\lsb\|\gtv^t\|_2^2\rsb+{\sf E}\lsb\|\nv_{e}^t\|_2^2\rsb\nonumber\\
&=&\f{1}{(n r^t)^2}{\sf E}\lsb\|\sum_{i\in[n]}\overline{\Uv}_{\rm r}\gv_i^t\|_2^2\rsb+d\sigma^2_{\nv_e}\nonumber\\
&=&\f{1}{(n r^t)^2}\sum_{j\in[d]}{\sf E}\lb\sum_{k\in[d]}\Uv_{{\rm r},j}^{\sf T}\Uv_{{\rm r},k}\lsb\sum_{i\in[n]}\gv_i^t\rsb_k\rb^2+d\sigma^2_{\nv_e}\nonumber\\
&=&\f{1}{(n r^t)^2}\sum_{j\in[d]}{\sf E}\lsb\left\|\Uv_{r,j}\right\|_2^4\rsb\lb\sum_{i\in[n]}\lsb\gv_i^t\rsb_j\rb^2
+\f{1}{(n r^t)^2}\sum_{j\in[d]}\sum_{k\in[d]}{\sf E}^2\lsb\lb\Uv_{{\rm r},j}^{\sf T}\Uv_{{\rm r},k}\rb^2\rsb\lb\sum_{i\in[n]}\lsb\gv_i^t\rsb_k\rb^2\nonumber\\
&&-\f{1}{(n r^t)^2}\sum_{j\in[d]}{\sf E}^2\lsb\lb\Uv_{{\rm r},j}^{\sf T}\Uv_{{\rm r},k}\rb^2\rsb\lb\sum_{i\in[n]}\lsb\gv_i^t\rsb_j\rb^2+d\sigma^2_{\nv_e}
\end{IEEEeqnarray}
where the $j^{\rm th}$ element of the equivalent noise vector $\nv_e^t$ is described as
\ea{\label{eq:jth-ne}
\lsb\nv_e^t\rsb_j =\f{1}{nc\sqrt{r^t}}\sum_{q\in[r]}\lsb\Uv_{r,j}\rsb_q\underbrace{\lb\sum_{i\in[n]}\sqrt{\f{\zeta_i^t \kappa_i^t}{r^t}}\lsb\mv_i^t\rsb_q+\lsb\nv^t\rsb_q\rb}_{\sim\Ncal\lb0,\sum_{i\in[n]}\f{\zeta_i^t P_i}{r}|h_i^t|^2 +\sigma^2_{\nv}\rb}.
}

Assuming the {\bf Rademacher distribution} for the entries of matrix $\Uv$ then $\nv_e^t\sim\Ncal\lb0,\sigma^2_{\nv_e}\Iv_d\rb$ where
\ea{\label{eq:equivalent-noise-variance}
\sigma^2_{\nv_e}=\f{1}{(n c^t)^2}\lb\sum_{i\in[n]}\f{\zeta_i \kappa_i^t}{r^t} +\sigma^2_{\nv}\rb.
}

Assuming the {\bf Achlioptas distribution} for the entries of matrix $\Uv$ then $\nv_e^t\sim 1/s\Ncal\lb0,s\sigma^2_{\nv_e}\Iv_d\rb+(1-1/s)\delta_{d}(\nv_e)$ where $\delta_{d}$ represents the deleta dirac function. It can be shown that for this distribution ${\sf E}\lsb\nv_{e}\nv_{e}^{\sf T}\rsb=\sigma^2_{\nv_e}\Iv_d$.

Assuming the {\bf Gaussian distribution} for the entries of matrix $\Uv$, then each term of the summation contributing to the $[\nv_e^t]_j$ has the following PDF
\ea{\label{eq:PDF-ne-GaussianU}
f(u_n)=\f{1}{\pi\sigma_{\nv_e}}K_0\lb\f{|u_n|}{\sigma_{\nv_e}}\rb
}
where $K_0(\ldotp)$ is the modified Bessel function of second type and order zero given as
\ea{\label{eq:modified-bessel-secondtype-zero-order}
K_0(z)=\int_{0}^{\infty}\cos{z\sinh{t}}dt=\int_0^\infty \f{\cos{tz}}{\sqrt{1+t^2}}dt.
}

However, we are not able to propose an explicit closed-form expression for the distribution of the equivalent noise unless we approximate it by a Gaussian distribution of zero mean and variance $\sigma^2_{\nv_e}$ in case of large reduced dimension and according to \emph{Central limit Theorem} (CLT).

Assuming the {\bf Rademacher distribution} for random matrix projection, then  $\Uv_{{\rm r},j}^{\sf T}\Uv_{{\rm r},k}$ is
\vspace{-2mm}
\ea{
\Uv_{{\rm r},j}^{\sf T}\Uv_{{\rm r},k}=\left\{
\begin{array}{cc}
   \|\Uv_{{\rm r},j}\|_2^2\stackrel{{\rm w.p.} 1}=r^t,  & k=j \\
    U_{\rm r}, & k\neq j
\end{array}
\right.
}
where $U_{\rm r}$ is a discrete r.v., with odd or even integer values between $-r^t$ and $r^t$ (depending on $r^t$ being odd or even), having zero mean and ${\sf E}[\lb\Uv_{{\rm r},j}^{\sf T}\Uv_{{\rm r},k}\rb^2]=r^t$. Hence, the second-order moment of the \emph{global} gradient estimation can be further simplified to
\begin{IEEEeqnarray}{rcl}\label{eq:convergence-Bernolli}
{\sf E}\lsb\|\hat{\gv}^t\|_2^2\rsb&=&\f{(r^t)^2 +r^t (d-1)}{(n r^t)^2}\|\sum_{i\in[n]}\gv_i^t\|_2^2+d\sigma^2_{\nv_e}\nonumber\\
&\stackrel{(b)}\leq&\frac{(r^t)^2 +r^t (d-1)}{(n r^t)^2}\lb\sum_{i\in[n]}\|\gv_i^t\|_2\rb^2+\f{d}{(n c^t)^2}\lb\sum_{i\in[n]}\f{\zeta_i \kappa_i^t}{r^t} +\sigma^2_{\nv}\rb\nonumber\\
&\stackrel{(c)}\leq& L^2\lb1+\f{d-1}{r^t}\rb+\f{d}{(n c^t)^2}\lb\sum_{i\in[n]}\f{\zeta_i \kappa_i^t}{r^t} +\sigma^2_{\nv}\rb.
\end{IEEEeqnarray}
where $(b)$ follows from the triangle inequality and $(c)$ follows by the $L$-smooth condition indicating that the loss function has $L$-Lipschitz continuous gradients and so $\|\gv_i^t\|_2\leq L$, $i\in[n]$.

Assuming the {\bf Achlioptas distribution} for sparse random projection then
\ea{
\Uv_{{\rm r},j}^{\sf T}\Uv_{{\rm r},k}=\left\{
\begin{array}{cc}
   \|\Uv_{{\rm r},j}\|_2^2,  & k=j \\
    U_{\rm r}, & k\neq j
\end{array}
\right.
}
where $\|\Uv_{{\rm r},j}\|_2^2/s\sim {\rm Bin}\lb r^t,1/s\rb$ and $U_{\rm r}$ is a discrete r.v., taking integer values between $-r^t$ and $r^t$ scaled by $s$, with zero mean and ${\sf E}\lsb\lb\Uv_{{\rm r},j}^{\sf T}\Uv_{{\rm r},k}\rb^2 \rsb=r^t$. Accordingly, second-order moment of the \emph{global} gradient estimation can be simplified as
\begin{IEEEeqnarray}{rcl}
{\sf E}\lsb\|\hat{\gv}^t\|_2^2\rsb&=&\f{1}{(n r^t)^2}\lsb r^t (s-1)+(r^t)^2+r^t (d-1)\rsb\|\sum_{i\in[n]}\gv_i^t\|_2^2+\sigma^2_{\nv_e}\nonumber\\
&\leq& L^2\lsb 1+\f{d+s-2)}{r^t}\rsb+\f{d}{(n c^t)^2}\lb\sum_{i\in[n]}\f{\zeta_i \kappa_i^t}{r^t} +\sigma^2_{\nv}\rb.
\end{IEEEeqnarray}

Assuming the {\bf Gaussian distribution} for random matrix projection then
\ea{
\Uv_{{\rm r},j}^{\sf T}\Uv_{{\rm r},k}=\left\{
\begin{array}{cc}
   \|\Uv_{{\rm r},j}\|_2^2\sim\chi^2(r^t),  & k=j \\
    U_{\rm r}, & k\neq j
\end{array}
\right.
}
where $U_{\rm r}$ is a continuous r.v. of zero mean and ${\sf E}\lsb\lb\Uv_{{\rm r},j}^{\sf T}\Uv_{{\rm r},k}\rb^2\rsb=r^t$, corresponding to the sum of $r^t$ independent r.v. each distributed according to \eqref{eq:PDF-ne-GaussianU} with zero mean and $\sigma^2=1$. The second-order moment of the \emph{global} gradient estimation can be simplified as
\begin{IEEEeqnarray}{rcl}
{\sf E}\lsb\|\hat{\gv}^t\|_2^2\rsb&=&\f{1}{(n r^t)^2}\lsb (r^t)^2+2r^t +r^t (d-1)\rsb\|\sum_{i\in[n]}\gv_i^t\|_2^2+\sigma^2_{\nv_e}\nonumber\\
&\leq&L^2\lsb\f{d+1}{r^t}+1\rsb+\f{d}{(n c^t)^2}\lb\sum_{i\in[n]}\f{\zeta_i\kappa_i^t}{r^t} +\sigma^2_{\nv}\rb.
\end{IEEEeqnarray}
\section{Proof of Th. \ref{th:optimal-convergence}}
\label{sec:appendix C}
The optimization problem under consideration can be considered equivalently as the following problem
\begin{IEEEeqnarray}{rcl}
&\min_{r,\{\zeta_i\}_{i\in[n]}}&  L^2\lb1+\f{d+s-2}{r}\rb+\f{d}{n^2 c^2}\lb\sum_{i\in[n]}\f{\zeta_i \kappa_i}{r} +1\rb\nonumber\\
&{\rm s.t.}&\ \f{\zeta_i\kappa_i}{r}\leq \f{\kappa_i}{r}(1-\gamma_i),\ \forall\ i\in[n]\nonumber\\
&&\sum_{i\in[n]}\f{\zeta_i \kappa_i}{r} +1\geq\max_{i\in[n]}\ (1+\varepsilon)\f{8\kappa_{\min} \ln\lb1.25/\delta\rb}
{(\epsilon_i^T /T)^2} \nonumber\\
&&r\geq (4+2a)\lb \varepsilon^{2}/2-\varepsilon^3/3\rb^{-1}\ln n.
\end{IEEEeqnarray}

For the objective function to be minimum, it suffices to first fix a value for $r$ starting at its lower bound in third constraint and then find the minimal set of clients satisfying the first two constraints. This requires finding the minimum number of users that can guarantee the $T$-fold privacy level through the accumulation of their noise powers at the PS. To this end, a similar approach to that of a water-filling scheme is leveraged as also used in \cite{FL_wireless_LDP}. More precisely, based on the power constraint, we sort clients in an decreasing order with respect to their rest of powers remaining from the alignment process which was designed so as to hold the unbiasedness condition. Then, we allocate $\zeta_i$ so as to satisfy the second constraint with equality releasing the following expression as the minimum of the noise power terms satisfying both constraints:
\begin{IEEEeqnarray}{rcl}
\omega_i =\f{\zeta_i \kappa_i}{r} =\min\lb\f{\kappa_i}{r}(1-\gamma_i),\lsb\Omega-\sum_{k=1}^{i-1}\f{\zeta_k \kappa_k}{r}\rsb^{+}\rb.
\end{IEEEeqnarray}
This in fact suggests that maximum level of noise power that can be supported by clients is allocated so long as their aggregation does not exceed the RHS of LDP constraint.
If the minimal set is empty, then the value of $r$ and the minimal set of clients obtained in previous iteration and their corresponding $\zeta_i$ coefficients yield the optimal solution of the problem. Otherwise, the process is repeated by  increasing $r$ as $r=r+1$ and find the next minimal set of clients.
\end{document}